\documentclass{webofc}
\usepackage[varg]{txfonts}   
\newcommand{\al}{&\!\!\!\!}
\newcommand{\lag}{\mathcal{L}}
\newcommand{\nno}{\nonumber\\}\newcommand{\be}{\begin{equation}}
\newcommand{\ee}{\end{equation}}
\newcommand{\bea}{\begin{eqnarray}}
\newcommand{\eea}{\end{eqnarray}}
\newcommand{\beas}{\begin{eqnarray*}}
\newcommand{\eeas}{\end{eqnarray*}}

\def\vec#1{\boldsymbol{#1}}

\newcommand{\pc}{P_c}
\newcommand{\jp}{J/\psi p}
\newcommand{\sigh}{\Sigma_c^{(*)}\bar{D}^{(*)}}

\newcommand{\lamh}{\Lambda_c\bar{D}^{(*)}}
\newcommand{\etacp}{\eta_c p}
\newcommand{\etacn}{\eta_c p}

\newcommand{\bonn}{Helmholtz-Institut f\"ur Strahlen- und Kernphysik and Bethe Center for Theoretical Physics,\\ Universit\"at Bonn, D-53115 Bonn, Germany}

\newcommand{\ific}{Instituto de F\'isica Corpuscular (centro mixto CSIC-UV),
Institutos de Investigaci\'on de Paterna,\\ Apartado 22085, 46071, Valencia, Spain
}

\begin{document}
\title{Molecular interpretation of the LHCb pentaquarks from an analysis of $\jp$ spectrum}
%
%

\author{\firstname{Meng-Lin} \lastname{Du}\inst{1,2}\fnsep\thanks{\email{du.menglin@ific.uv.es}} 
}

\institute{\ific 
\and
           \bonn 
          }

\abstract{A coupled-channel approach including the $\lamh$ and $\etacp$ channels in addition to the $\sigh$ and $\jp$ channels, as required by unitarity and heavy quark spin symmetry (HQSS) obtain cutoff independent results, the one-pion exchange potential in the multichannel systems is to be supplemented with next-leading order counter terms responsible for the $S$-wave-to-$D$-wave transitions. We show that the experimental data for the $\jp$ mass distributions are fully in line with the $\Sigma_c\bar{D}$ and $\Sigma_c\bar{D}^\ast$ hadronic molecular interpretation of the $\pc(4312)$ and $\pc(4440)/\pc(4457)$, respectively. A narrow $\Sigma_c^\ast\bar{D}$ molecule around 4.38 GeV is required by the HQSS with the evidence for its existence seen in the $\jp$ spectrum. Moreover, we predict the line shapes for the elastic and inelastic channels.}
\maketitle
\section{Introduction}
\label{intro}
The observation of the $\pc(4380)$ and $\pc(4450)$ \cite{LHCb:2015yax}, as the first well established signals of pentaquarks, in 2015 builds a landmark in searching for the so-called exotic states allowed by quantum chromodynamics (QCD), i.e., the color neutral objects composed rather than a pair of quark and antiquark ($q\bar q$) or three quarks ($qqq$). 
These exotic structures are observed in the $\jp$ channel in $\Lambda_b^0\to J/\psi K^- p$ decays, which  implies that they contain at least five quarks, $c\bar cuud$. The $\pc(4450)$ is confirmed by an updated analysis based on an combined data set collected in Run 1 and Run 2 by LHCb Collaboration in 2019 \cite{LHCb:2019kea}, and it is observed to consist of two narrow overlapping structures, $\pc(4440)$ and $\pc(4457)$. Nevertheless, since the one-dimensional fits to the $m_{\jp}$ distributions alone cannot distinguish broad $\pc$ structures from background varying slowly with $m_{\jp}$, a verification of the broad $\pc(4380)$ observed in Ref.~\cite{LHCb:2015yax} awaits an full amplitude analysis of the new data. At the same time, a new narrow pentaquark state $\pc(4312)$ is discovered in the same channel \cite{LHCb:2019kea}. The discovery of these pentaquark states immediately spurred many theoretical works and various interpretations on the nature of these pentaquark states were proposed, including hadronic molecules~\cite{Chen:2019bip,Chen:2019asm,Liu:2019tjn,Guo:2019kdc,Xiao:2019aya,Meng:2019ilv,Du:2019pij,Du:2021fmf}, compact pentaquark states~\cite{Cheng:2019obk,Ali:2019npk,Zhu:2019iwm}, hadrocharmonia~\cite{Eides:2019tgv}, and cups effects~\cite{Kuang:2020bnk,Nakamura:2021qvy}. An amplitude analysis focusing on the $\pc(4312)$ peak suggests the virtual state origin of the $\pc(4312)$ in Ref.~\cite{Fernandez-Ramirez:2019koa}. Among these explanations, the molecular picture is of particular interest since all of the three narrow pentaquark states can be simultaneously interpreted as $\Sigma_c\bar{D}^{(*)}$ bound states with help of heavy quark spin symmetry (HQSS). The charmonium pentaquark with strangeness was predicted in Refs.~\cite{Wu:2010jy,Chen:2016ryt,Xiao:2019gjd,Wang:2019nvm} and its signal was first observed by LHCb Collaboration in 2020 \cite{LHCb:2020jpq}, i.e., $P_{cs}(4459)$. Its observation has promoted numerous studies of this resonance~\cite{Peng:2020hql,Chen:2020uif,Lu:2021irg,Du:2021bgb}. These studies typically favor a molecular picture with a dominant component of $\Xi_c\bar{D}^*$.

The proximity of the $\Sigma_c\bar{D}^{(*)}$ thresholds to these narrow $\pc$ suggests an important role played by the corresponding two-hadron states in the dynamics of the $\pc$ states. In the common molecular picture, the $\pc(4312)$ is regarded as an $S$-wave $\Sigma_c\bar{D}$ bound state, while the $\pc(4440)$ and $\pc(4457)$ are bound states of $\Sigma_c\bar{D}^*$ with different spin structures. In order to describe the measured $\jp$ mass distributions, the inelastic channels need to be considered to account for the width of $\pc$'s. In addition to the $\jp$ channel, there are more inelastic channels, most prominently $\lamh$ and $\etacn$. The latter is connected to the $\jp$ channel via HQSS. When the $\lamh$ channel is included explicitly, the three-body cut $\Lambda\bar{D}^\ast\pi$ effects need to be taken into account as the width of $\Sigma_c^{(*)}$ is almost saturated by the decay $\Sigma_c^{(*)}\to \Lambda_c\pi$. In this talk, we will include all the elastic channels $\sigh$ and inelastic channels, $\jp$, $\etacn$ and $\lamh$, explicitly, and investigate the role played by the $\lamh$ and the nonperturbative dynamical pions as the proximity of the $\pc$ states to the $\lamh\pi$ thresholds. For the case of only short-distance contact potentials, two solutions were found describing the data almost equally well~\cite{Du:2019pij}. The most prominent contribution from the one-pion exchange (OPE) originates from the $D$-waves, especially the $S$-wave-to-$D$-wave transitions \cite{Baru:2016iwj,Baru:2017gwo,Wang:2018jlv}. Renormalization of the OPE in the energy region of interest requires an inclusion of higher order contact potentials \cite{Baru:2017gwo,Wang:2018jlv}. When the OPE potentials are included together with the necessary next-leading order (NLO) $S$-$D$ counter terms, only one of the two solutions present in purely contact potentials survives under the requirement of the renormalizability (regulator-independent), which suggests that the quantum numbers of the $\pc(4440)$ and $\pc(4457)$ should be $J^P=\frac{3}{2}^-$ and $\frac12^-$, respectively. We also show that there is evidence for the existence of a narrow $\Sigma_c^*\bar{D}$ bound state, required by the HQSS, in the $m_{\jp}$ mass distribution around 4.38~GeV. The width of this $\pc(4380)$ is of the same order of the three narrow $\pc$ states as a consequence of HQSS and thus different from the broad one reported by LHCb in 2015 \cite{LHCb:2015yax}.

\section{Framework}
\label{sec-1}

\subsection{Interactions}
\label{sec:Lag}

The effective interactions contain both short-range contact potentials and long-range OPE. The leading-order (LO) effective Lagrangian for the $\sigh$ and $\lamh$ interactions respecting HQSS reads 
\bea\label{lag:contactelastic}
\lag_\text{LO} \al =\al - C_a S^{i\dag }_{ab} S_{ba}^i \langle \bar{H}_c^\dag\bar{H}_c\rangle - C_b i\epsilon_{jik} S_{ab}^{j\dag}S_{ba}^k \langle \bar{H}_c^\dag \sigma^i\bar{H}_c\rangle \nno
\al\al  + C_c \Big(S_{ab}^{i\dag} T_{ca}\langle \bar{H}_c^\dag\sigma^i \bar{H}_b\rangle - T^\dag_{ca}S_{ab}^i\langle \bar{H}^\dag_b\sigma^i\bar{H}_c\rangle\Big) - C_d T^\dag_{ab}T_{ba}\langle \bar{H}_c^\dag \bar{H}_c\rangle,
\eea
where $\langle\ldots\rangle$ represents the trace in the spinor space and $\sigma$ denotes the Pauli matrices. The subindices $a$, $b$, $c$ are the flavor indices, and the superindices $i$, $j$, $k$ denote the polarization. The heavy quark spin doublets for the ground states  $(\Sigma_c,\Sigma_c^*)$, $(\bar{D},\bar{D}^*)$, and $\Lambda_c^+$ are grouped in $S^i$, $\bar{H}$, and $T$, respectively, ~\cite{Manohar:2000dt}
\bea
S^i=\frac{1}{\sqrt{3}}\sigma^i\Sigma_c + \Sigma^{*i}_c,\qquad \bar{H}=-\bar{D}+\sigma\cdot \bar{D}^*,
\eea
with 
\bea
S_c^{(*)}=\begin{pmatrix}
\Sigma_c^{(*)++} & \frac{1}{\sqrt{2}}\Sigma_c^{(*)+} \\
  \frac{1}{\sqrt{2}}\Sigma_c^{(*)+} & \Sigma_c^{(*)0} \\
\end{pmatrix}, \quad
 \quad
\bar{D}^{(*)} = \begin{pmatrix}
\bar{D}^{(*)0} \\
D^{(*)-} \\
\end{pmatrix}, \quad
T=\begin{pmatrix}
0 & \frac{1}{\sqrt{2}}\Lambda_c^+ \\
- \frac{1}{\sqrt{2}}\Lambda_c^+& 0\\
\end{pmatrix}. 
\eea
The transitions between the $\sigh$ and $\jp$ ($\etacn$) can be described by~\cite{Sakai:2019qph}:
\bea
\lag_\text{ine} = \frac{g_S}{\sqrt{3}}N^\dag \sigma^i \bar{H} J^\dag S^i - \sqrt{3}g_DN^\dag \sigma^i\bar{H}(\partial^i\partial^j-\frac13\delta^{ij}\partial^2)J^\dag S^j,
\eea
where $N$ denotes the nucleon doublet, and $J=-\eta_c+\vec\sigma\cdot \vec\psi$ collects the ground state charmonium fields. The LO contact potentials can be derived from a more intuitive way by decomposing the spin structures into the heavy and light degrees of freedom, see e.g. in Refs.~\cite{Du:2019pij,Du:2021fmf}. 

The LO OPE potential can be obtained by the effective Lagrangian for the axial coupling of pions to charmed mesons and baryons~\cite{Yan:1992gz}, 
\bea\label{lag:ope}
\lag = \frac{g_1}{4} \langle \vec{\sigma}\cdot \vec{u}_{ab}\bar{H}_b\bar{H}_a^\dag \rangle + i g_2\epsilon_{ijk} S^{i\dag}_{ab}u^j_{bc}S^k_{ca} - \frac{1}{\sqrt{2}}g_3\big( S_{ab}^{i\dag}u_{bc}^i T_{ca} + T^\dag_{ab} u_{bc}^i S^{i}_{ca} \big),
\eea
where $\vec{u} = -\nabla \Phi/F_\pi$ with $\Phi=\vec{\tau}\cdot\vec{\pi}$. Here $F_\pi=92.1$ MeV is the pion decay constant and $g_1=0.57$ is determined from the width of $D^{*+}\to D^0\pi^+$, $g_2=0.42$ and $g_3=0.71$ are taken from a lattice QCD calculation~\cite{Detmold:2012ge}. 

\subsection{Lippmann-Schwinger equations}\label{sec:lse}

In order to describe the $\Lambda_b^0\to \jp K^-$ decay, we need the bare weak production vertices as well. In this work, we only consider the $S$-wave $\sigh$ source for the $\Lambda_b^0\to \sigh K^-$ process since the $\sigh$ thresholds are close to the energy of interest. We introduce seven parameters for the all seven $S$-wave $\sigh$ sources \cite{Du:2019pij,Du:2021fmf}. 

\begin{figure}[htb]
\centering
\includegraphics[width=0.5\textwidth]{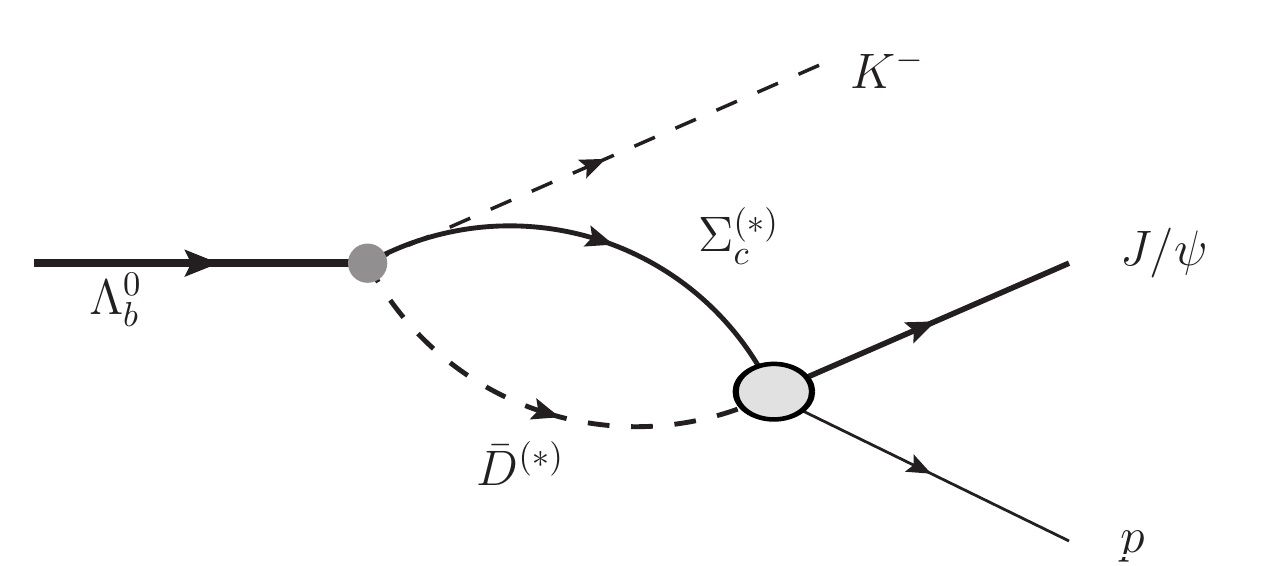}
\caption{Diagram for the $\Lambda_b^0\to\jp K^-$ through the $\Lambda_b^0\to\sigh K^-$.}
\label{fig-1}       
\end{figure}

With the above ingredients, one can obtain the production amplitude satisfying unitarity by solving the Lippmann-Schwinger equations (LSEs), with a diagrammatic representation shown in Fig.~\ref{fig-1},
\bea\label{eq:lse}
U^J_\alpha(E,p) \al = \al  P^J_\alpha {-}\! \sum_\beta \!\! \int \!\! \frac{d^3\vec{q}}{(2\pi)^3}V^J_{\alpha\beta}(E,p,q)G_\beta(E,q) U^J_\beta(E,q),  \nno
U^{(\prime) J}_i(E,k)\al =\al  {-}\! \sum_\beta \!\! \int \!\! \frac{d^3 \vec{q}}{(2\pi)^3} {V}^{(\prime)J}_{i\beta}(k) G_\beta(E,q) U^J_\beta(E,q),
\eea
with $U_\alpha^J$ and $P_\alpha$ denote the production amplitude and the source for the $\alpha^\text{th}$ elastic channel ($\lamh$ channels as well when they are included) and $U_i^{(\prime) J}$ for the $i^\text{th}$ inelastic channel $\jp$ ($\etacn$). The direct $\jp$ and $\etacn$ interactions are neglected since they are Okubo-Zweig-Iizuka suppressed and were found very weak in a recent lattice QCD calculation~\cite{Skerbis:2018lew} and a coupled-channel model in Ref.~\cite{Du:2020bqj}. In addition, we assume that the effect from the interactions of the $\lamh$ with $\jp$ and $\etacp$ can be neglected.

To take into account the widths of $\Sigma_c^{(*)}$ which are comparable to those of the $\pc$ states, the two-body propagator is taken as the following form
\bea\label{eq:twoprop}
G_\beta (E,\vec{q}) = \frac{m_{\Sigma_c^{(*)}}m_{D^{(*)}}}{E_{\Sigma_c^{(*)}}(\vec{q})E_{D^{(*)}}(\vec{q})}\frac{1}{E_{\Sigma_c^{(*)}}(\vec{q})+E_{D^{(*)}}(\vec{q})-E-  {\tilde{\Sigma}_R(s,m_{\Sigma_c^{(*)}})}/{(2E_{\Sigma_c^{(*)}}(\vec{q})})},
\eea
with $s=\big (E-E_{D^{(*)}}(\vec{q}) \big)^2 - \vec{q}^2$ for the off-shell $\Sigma_c^{(*)}$ and $\tilde{\Sigma}_R(s,m_{\Sigma_c^{(*)}} ) = i g^2{\lambda^{3/2}(s,m_{\Lambda_c}^2,m_\pi^2)}/{(8s^{5/2})}$ the renormalized self-energy loop diagram. The effect of the $\bar{D}^{(*)}$ width is neglected as it is only several tens of keV. To render the integrals in the LSEs in Eq.~\eqref{eq:lse} well defined we introduce a regulator, with a hard cutoff $\Lambda$ larger than all typical three-momenta scales to the relevant degrees of freedom.

\section{Fit results}\label{sec-2}

To fit the $\jp$ invariant mass distribution, a smooth incoherent background is exploited to model the contributions from misidentified non-$\Lambda_b^0$ events, crossed-channel effects and possibly additional broad $\pc^+$ structures \cite{Du:2021fmf}. First we perform fits employing only the contact potentials, referred to as Scheme I. In this circumstance, it is found that the data can be well described without including the $\lamh$ channels and the inclusion of these channels does not improve the fit quality. As in Ref.~\cite{Du:2019pij}, two solutions, denoted as $A$ and $B$, are found describing the data almost equally well. Both solutions give seven poles for the $\sigh$ scattering amplitudes, i.e., seven $\pc$ states. In both solutions, the $\pc(4312)$ corresponds to a $\Sigma_c\bar{D}$ bound state with $J^P=\frac12^-$. The $\pc(4440)$ and $\pc(4457)$ are bounds states of $\Sigma_c\bar{D}^*$, with quantum numbers $\frac12^-$ and $\frac32^-$, respectively, in solution $A$ and however interchanged in solution~$B$. Three $\pc$ states dominated by the $\Sigma_c^*\bar{D}^*$ are found in two solutions with $m_{\frac12^-}<m_{\frac32^-}<m_{\frac52^-}$ for solution $A$ and the opposite mass pattern for $B$. Those states are invisible in the $m_{\jp}$ mass distributions and were suggested to search for in a prompt production at the center-of-mass energy 7 TeV in the $pp$ collision \cite{Ling:2021sld}. In particular, in both solutions there is a narrow pole around 4.38 GeV corresponding to a $\Sigma_c^*\bar{D}$ bound state with $J^P=\frac32^-$. It is a consequence of the HQSS and different to the broad resonance reported by LHCb in 2015~\cite{LHCb:2015yax}.

\begin{figure*}[htb]
 \centering
  \includegraphics[width=0.32\textwidth]{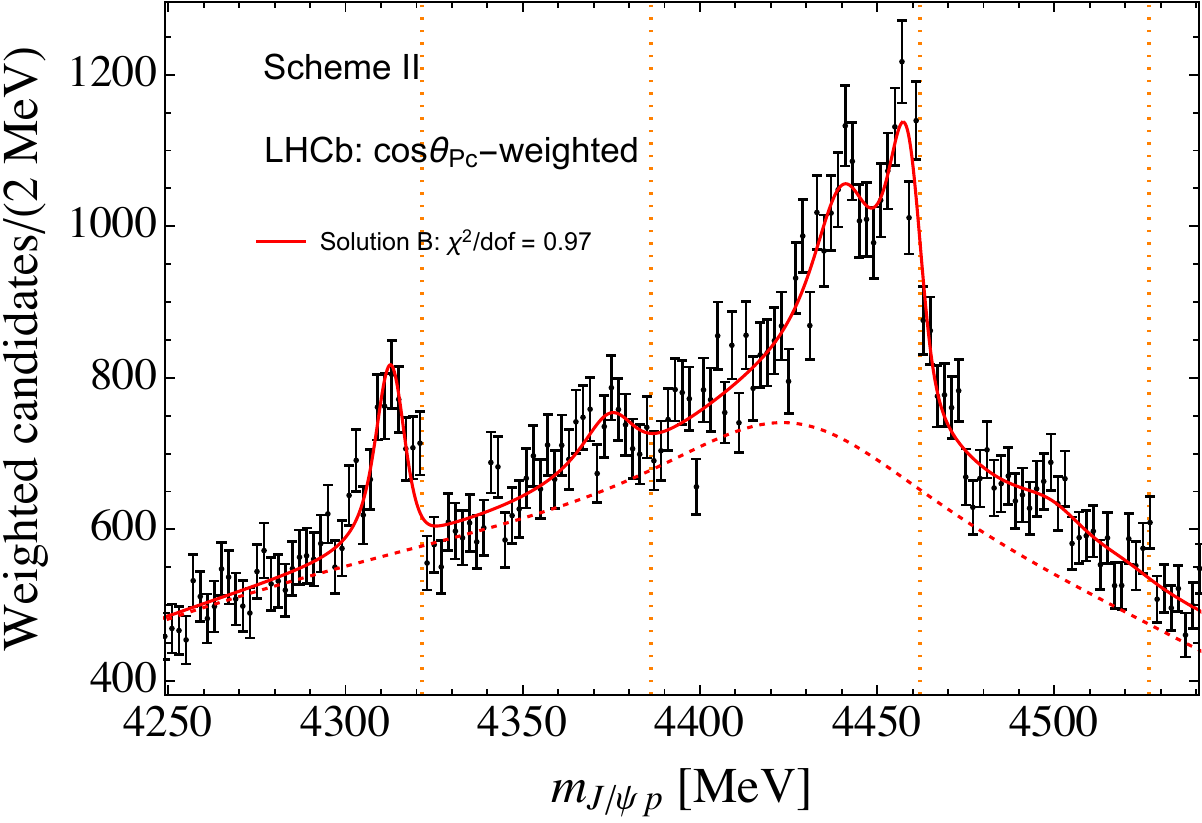}
  \includegraphics[width=0.32\textwidth]{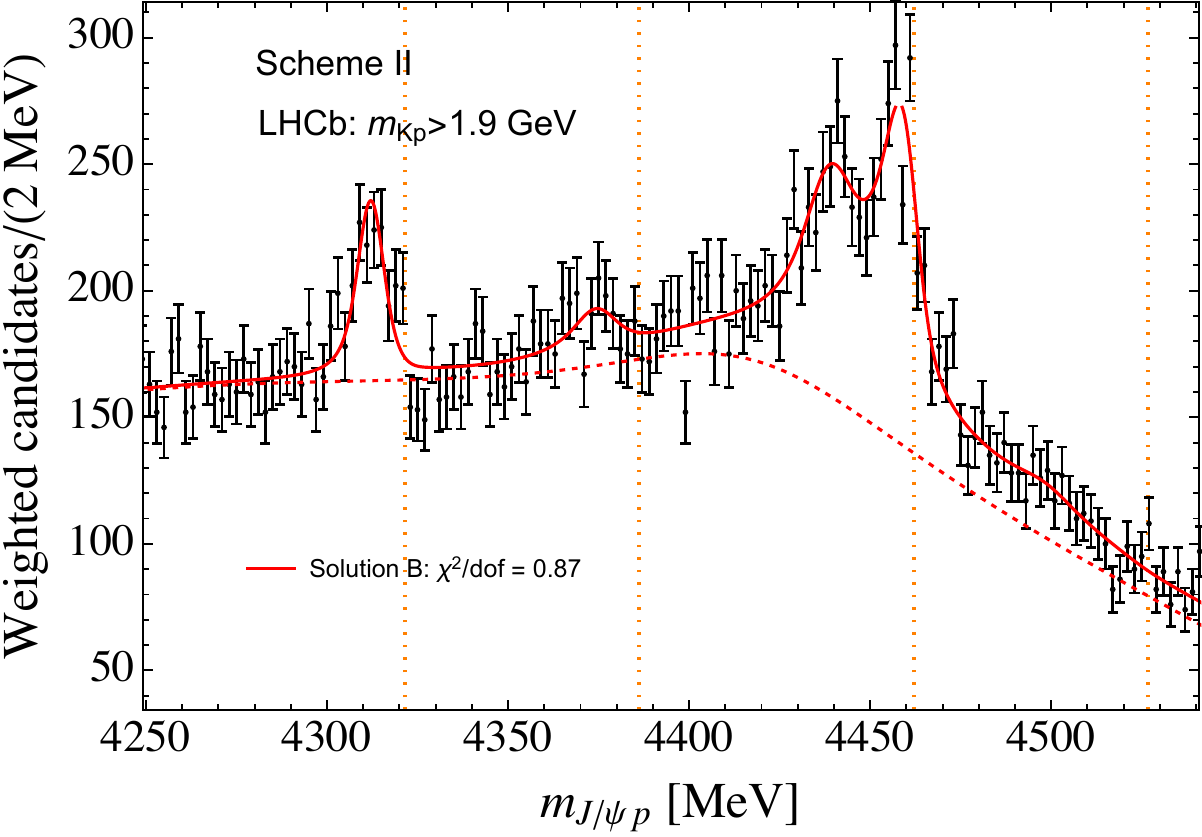}
  \includegraphics[width=0.32\textwidth]{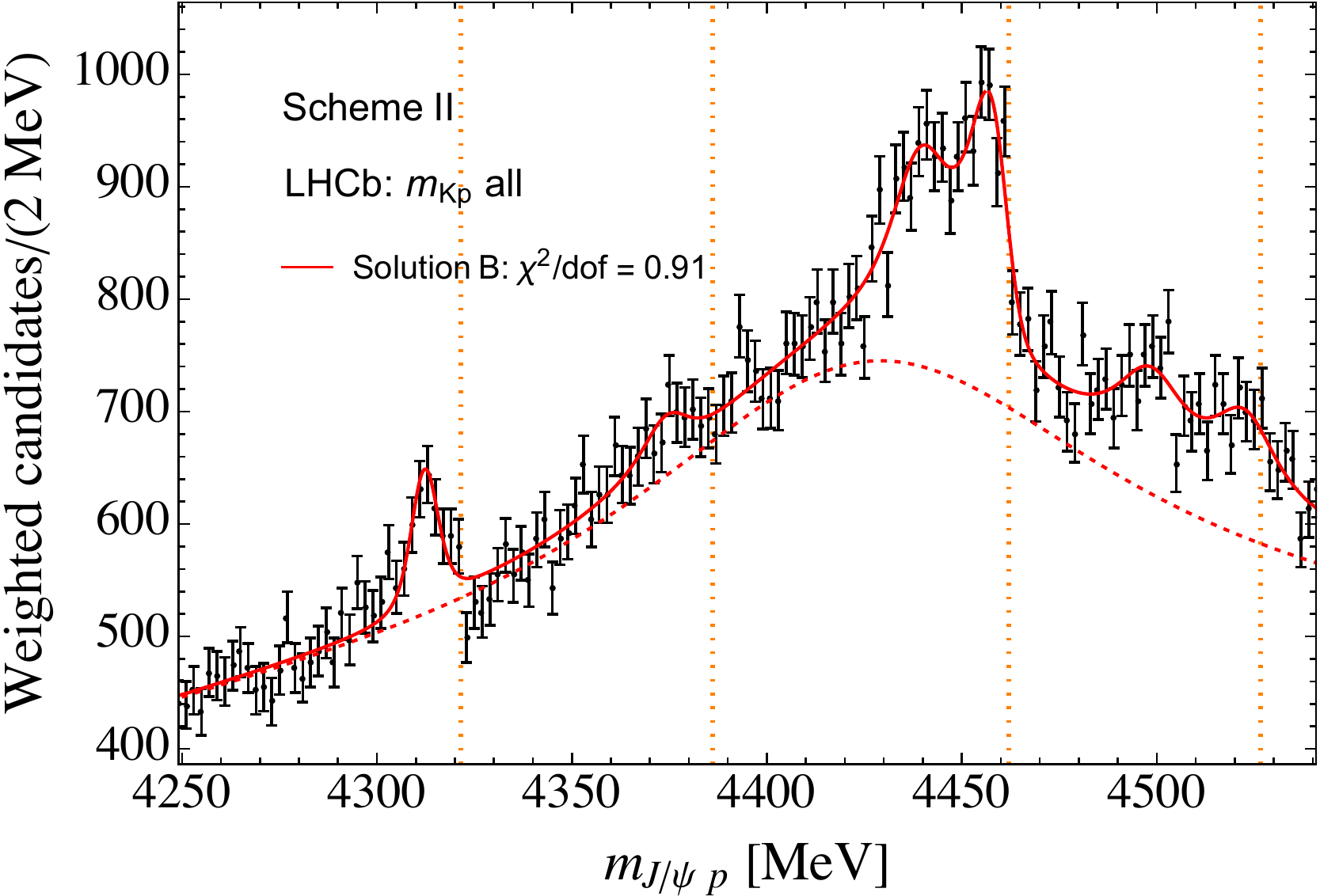}\\
  \includegraphics[width=0.33\textwidth]{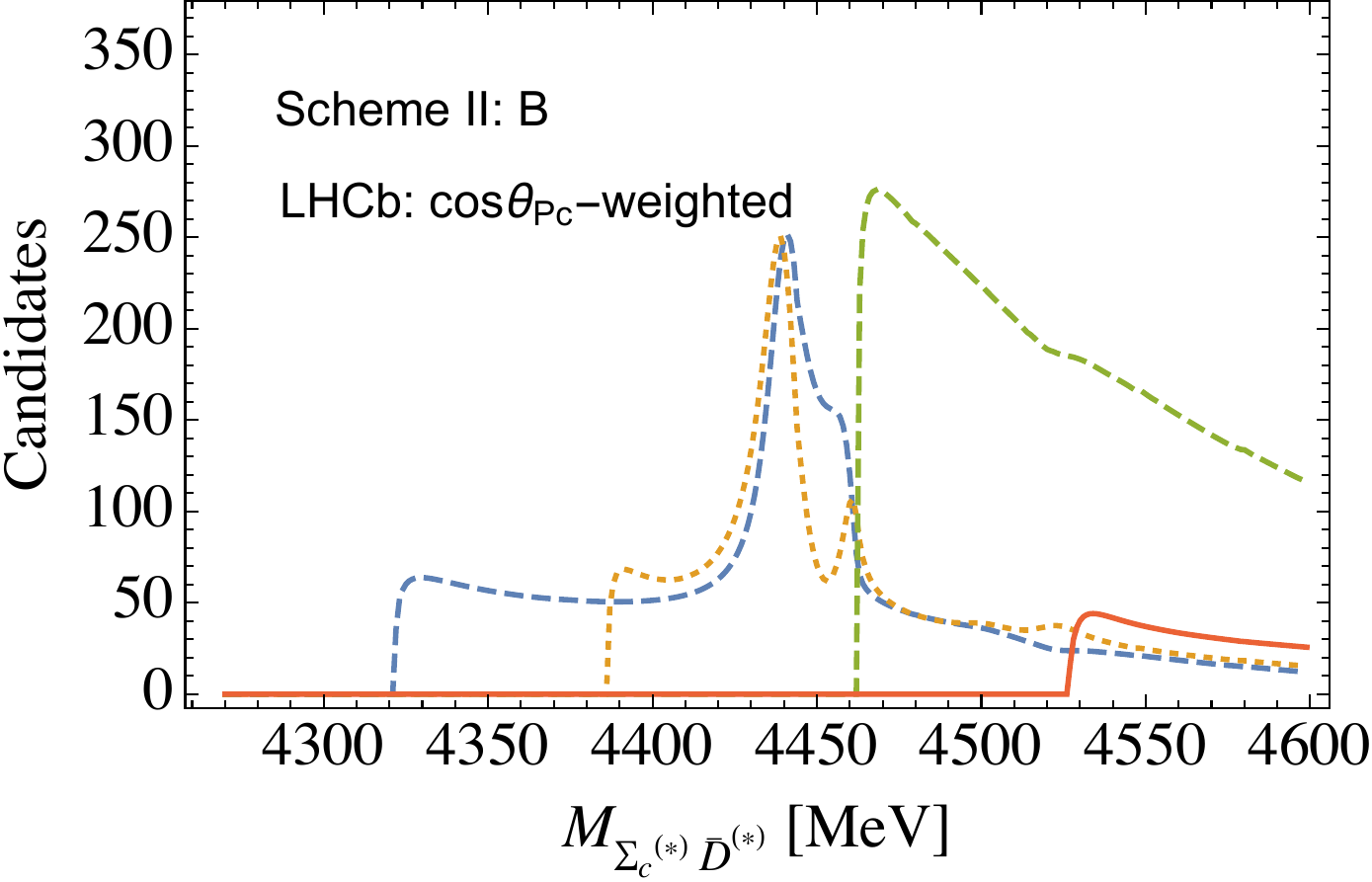}\includegraphics[width=0.33\textwidth]{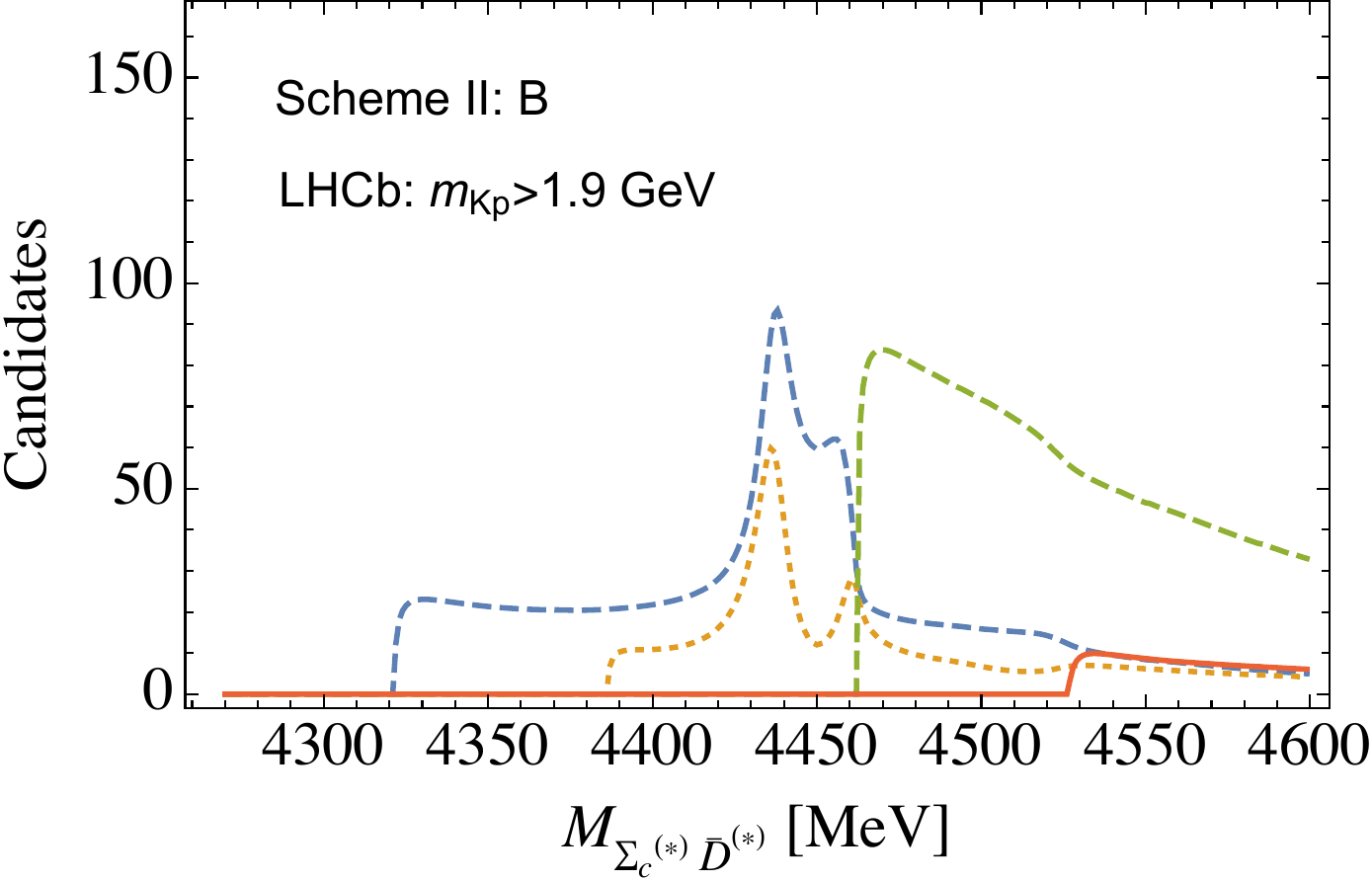}
  \includegraphics[width=0.33\textwidth]{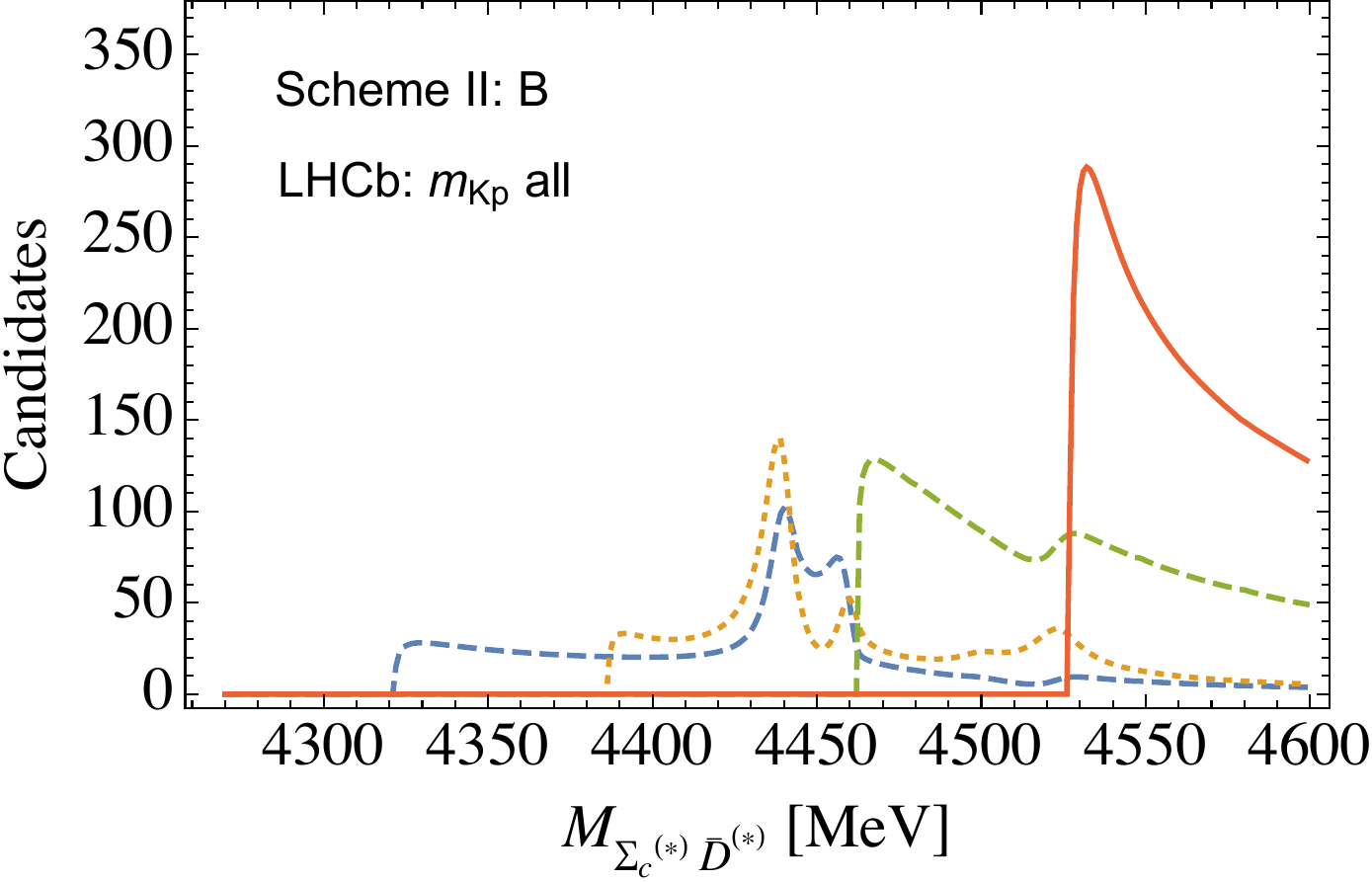}
  \caption{First row: The best fits to the different experimental data~\cite{LHCb:2019kea} for solution $B$ (red solid curves). The corresponding backgrounds are shown as red-dotted  curves. Second row: The corresponding predictions for the line shapes of $\Sigma_c\bar{D}$ (blue dashed), $\Sigma_c^\ast \bar{D}$ (orange dotted), $\Sigma_c\bar{D}^\ast$ (green dot-dashed), and $\Sigma_c^\ast\bar{D}^\ast$ (red solid) mass distributions.}
  \label{fig:fits:scheme_II}
\end{figure*}

As inspired by Scheme I, we assume that the widths of the $\pc$ states are saturated by the $\jp$, $\etacp$, the elastic channels (including those via the widths of $\Sigma_c^{(*)}$) and investigate the effects from the OPE for the elastic channels. The importance of the OPE is known as its tensor force mixes the $S$ and $D$ waves and can leave a significant impact on the line shapes. The large typical momenta, $p_\text{typ}\sim\sqrt{2\mu\delta}\sim 670$~MeV with $\delta$ and $\mu$ the largest threshold splitting and the reduced mass of the $\sigh$ system, make the contribution from the tensor force of the OPE even more important than that in the nucleon interaction \cite{Baru:2016iwj,Baru:2017gwo}. The iteration of such a potential within the LSEs was shown to yield a strong regulator dependence for the observable quantities. The cutoff-dependence for the $\jp$ and the $\sigh$ invariant mass distributions are demonstrated in Ref.~\cite{Du:2021fmf}. In order to cure the cutoff dependence, we include the NLO $S$-$D$ contact terms for the elastic channels together with the OPE as suggested in Refs.~\cite{Wang:2018jlv}. In what follows, this formulation will be referred to as Scheme II. In this case, there are in principle two best fits corresponding $A$ and $B$ for small cutoffs. The solution $B$ demonstrate only a milder regulator dependence, especially for cutoffs $\Lambda>1$~GeV which provide a larger separation between the soft and hard scales. On the contrary, the solution $A$ still show a strong cutoff dependence and the $\jp$ line shape can not be well described with cutoffs greater than 1.5~GeV, especially for the peak $\pc(4440)$. Therefore, we discard the solution $A$ and only focus on the results of solution $B$. More details on the cutoff dependence for two solutions can be found in Ref.~\cite{Du:2021fmf}.

\begin{figure*}[tb]
 \centering
  \includegraphics[width=0.45\textwidth]{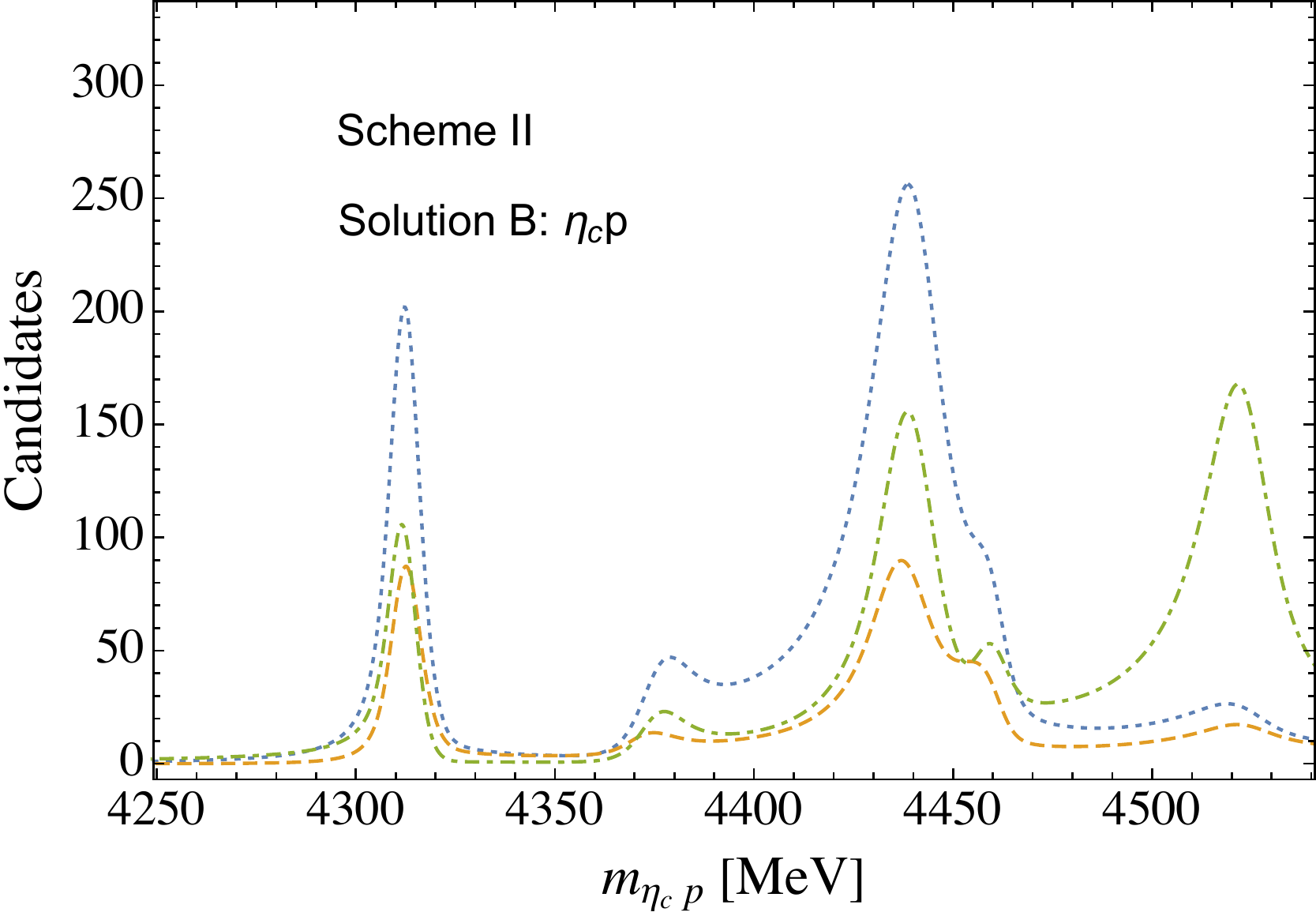}
  \includegraphics[width=0.45\textwidth]{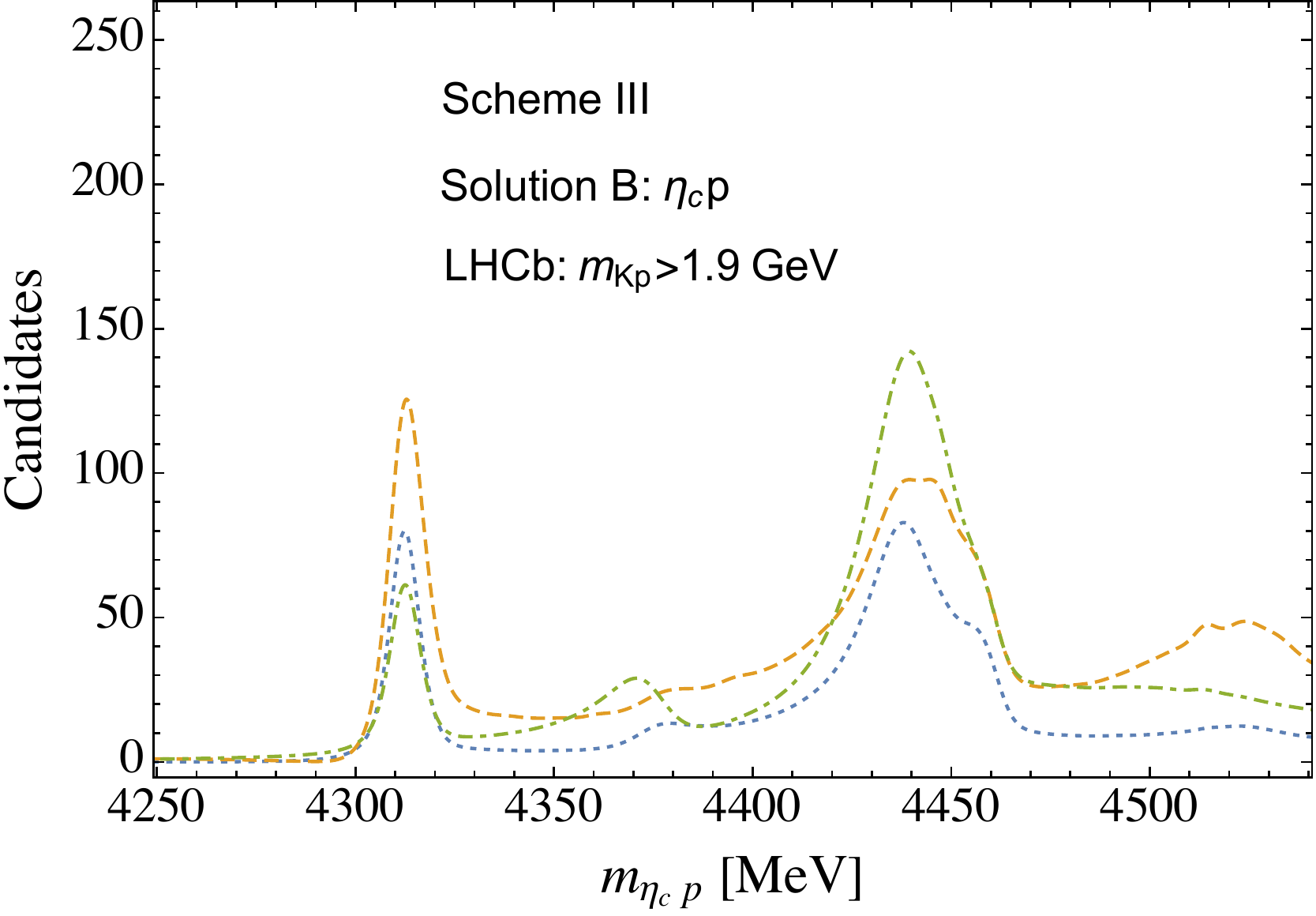}
  \caption{ \label{fig:etacn} 
 Predictions for the line shapes of the $\etacn$ invariant mass distributions based on  various fit results for  solution $B$ of Scheme~II (left) and Scheme~III (right), respectively. The blue dotted, orange dashed and  green dot-dashed curves correspond to the predictions from the three fit results ($\cos\theta_{P_c}$-weighted, $m_{Kp}>1.9$ GeV and $m_{Kp}$-all), in order. }
\end{figure*}

The best fits to the different experimental data \cite{LHCb:2019kea} are shown in Fig.~\ref{fig:fits:scheme_II} with corresponding predictions for the line shapes of the $\sigh$ mass distributions. Predictions for the $\etacn$ line shape are shown in Fig.~\ref{fig:etacn}. 
The poles lies in the complex energy plane of unphysical Riemann sheets associated to the amplitudes are identified as the $\pc$ resonances, which are collected in Table~\ref{tab-1} for those of interest. The $\pc(4312)$ couples dominantly to $\Sigma_c\bar{D}$ with $J^P=\frac12^-$ as a (quasi) bound state. The $\pc(4440)$ and $\pc(4457)$ couple dominantly to the $\Sigma_c\bar{D}^\ast$ with $J^P=\frac32^-$ and $J^P=\frac12^-$, respectively. In particular, there is a narrow state located around 4.38 GeV where the data show a peak, though less prominent than the well-known three pentaquark states.

\begin{table}[htb]
\centering
\caption{The pole positions (on the sheets close to the physical region), quantum numbers, the dominant channels (DCs) and their thresholds.}
\label{tab-1}       
\begin{tabular}{l|ccc}
\hline
\, & $J^P$ & DC (threshold [MV]) & Pole [MeV]  \\\hline
$\pc(4312)$ & $\frac12^-$ & $\Sigma_c\bar{D}$ (4321.6) & $4313(1)-3(1)i$ \\
$\pc(4380)$ & $\frac32^-$ & $\Sigma_c^\ast\bar{D}$ (4386.2) & $4376(1)-6(2)i$ \\
$\pc(4440)$ & $\frac{3}{2}^-$ & $\Sigma_c\bar{D}^\ast$ (4462.1) & $4441(2)-6(2)i$ \\
$\pc(4457)$ & $\frac12^-$ & $\Sigma_c\bar{D}^\ast$ (4462.1) & $4461(2)-5(2)i$ \\
$\pc$ & $\frac12^-$ & $\Sigma_c^\ast\bar{D}^\ast$ (4526.7) & $4525(4)-9(1)i$ \\
$\pc$ & $\frac32^-$ & $\Sigma_c^\ast\bar{D}^\ast$ (4526.7) & $4520(3)-12(3)i$ \\
$\pc$ & $\frac12^-$ & $\Sigma_c^\ast\bar{D}^\ast$ (4526.7) & $4500(2)-9(6)i$\\
\hline
\end{tabular}
\end{table}

\begin{figure*}[htb]
 \centering
  \includegraphics[width=0.5\textwidth]{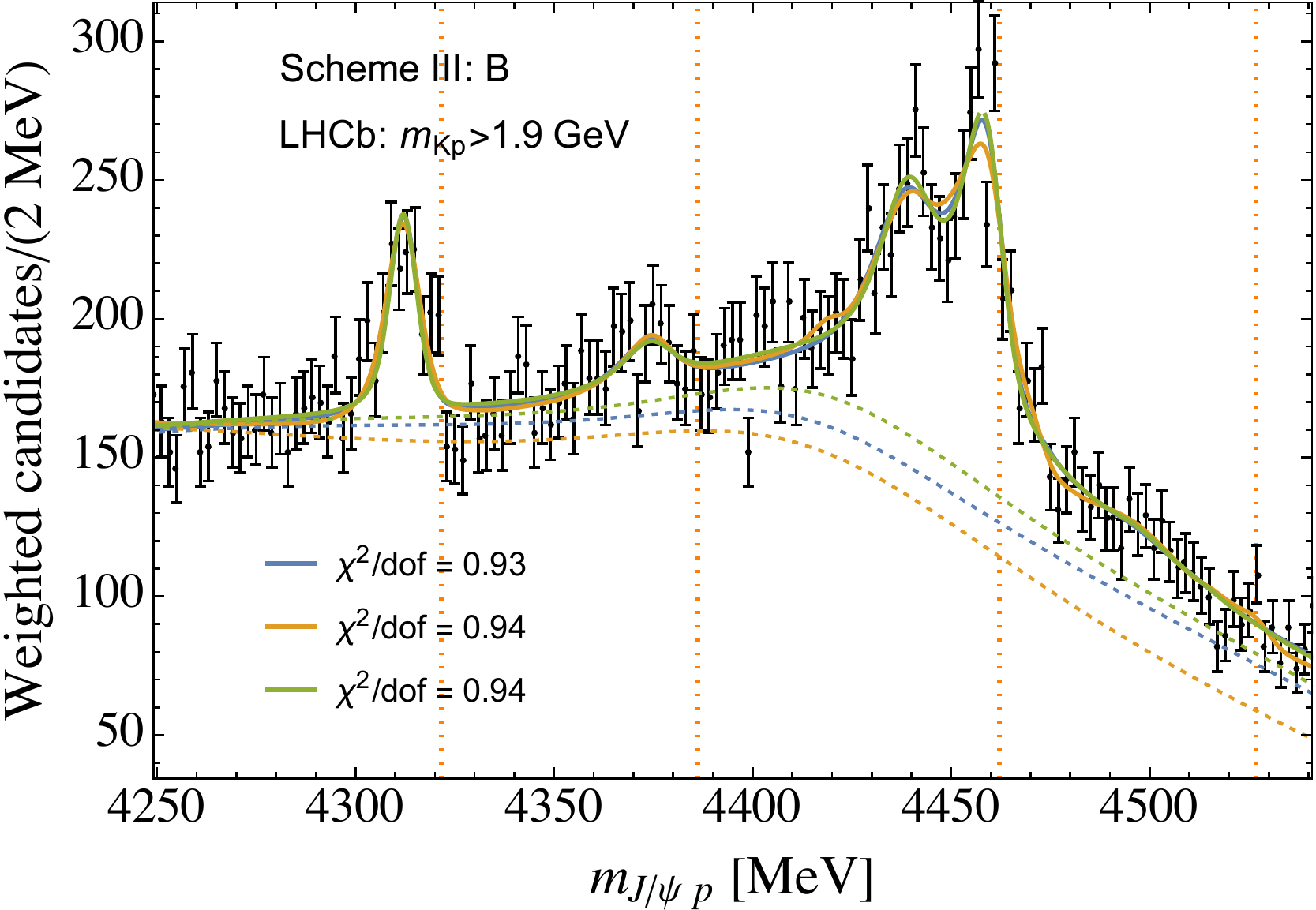}
  \caption{The best fits to the experimental data of $m_{Kp}>1.9$~GeV~\cite{LHCb:2019kea} with three different fixed backgrounds for Scheme~III.}
  \label{fig:III:fits}
\end{figure*}

\begin{figure*}[tb]
 \centering
 \includegraphics[width=0.32\textwidth]{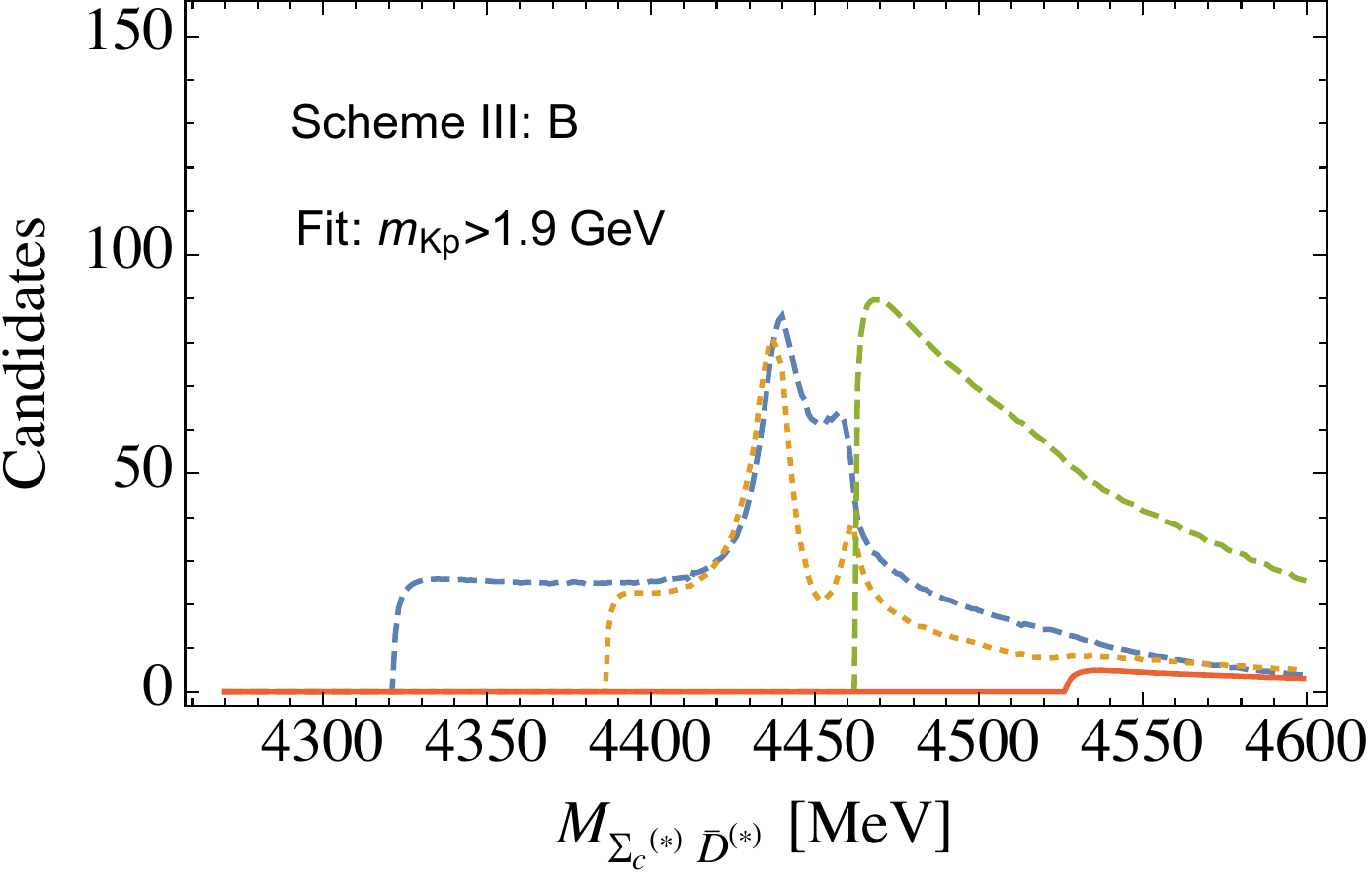}
   \includegraphics[width=0.34\textwidth]{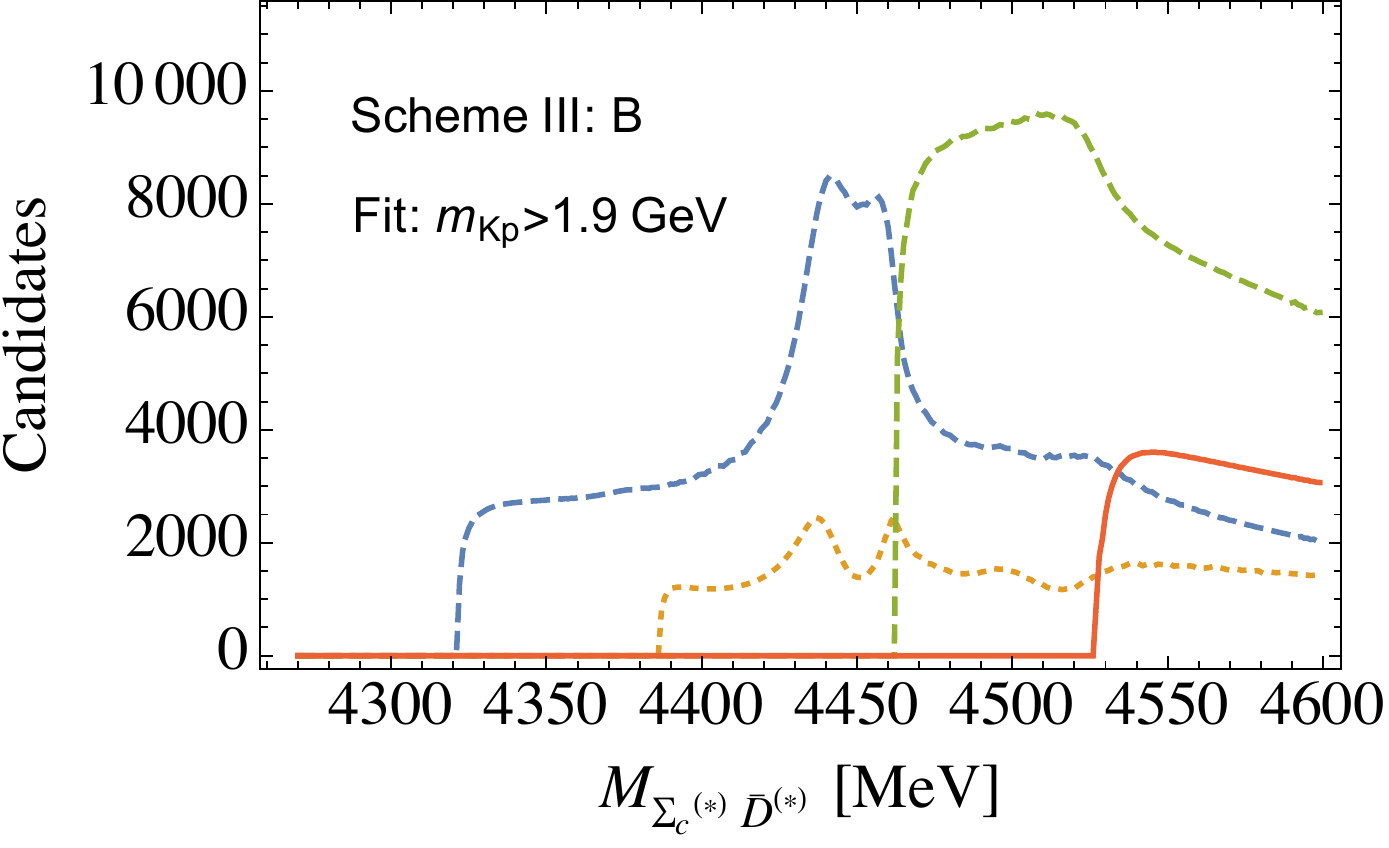} 
  \includegraphics[width=0.32\textwidth]{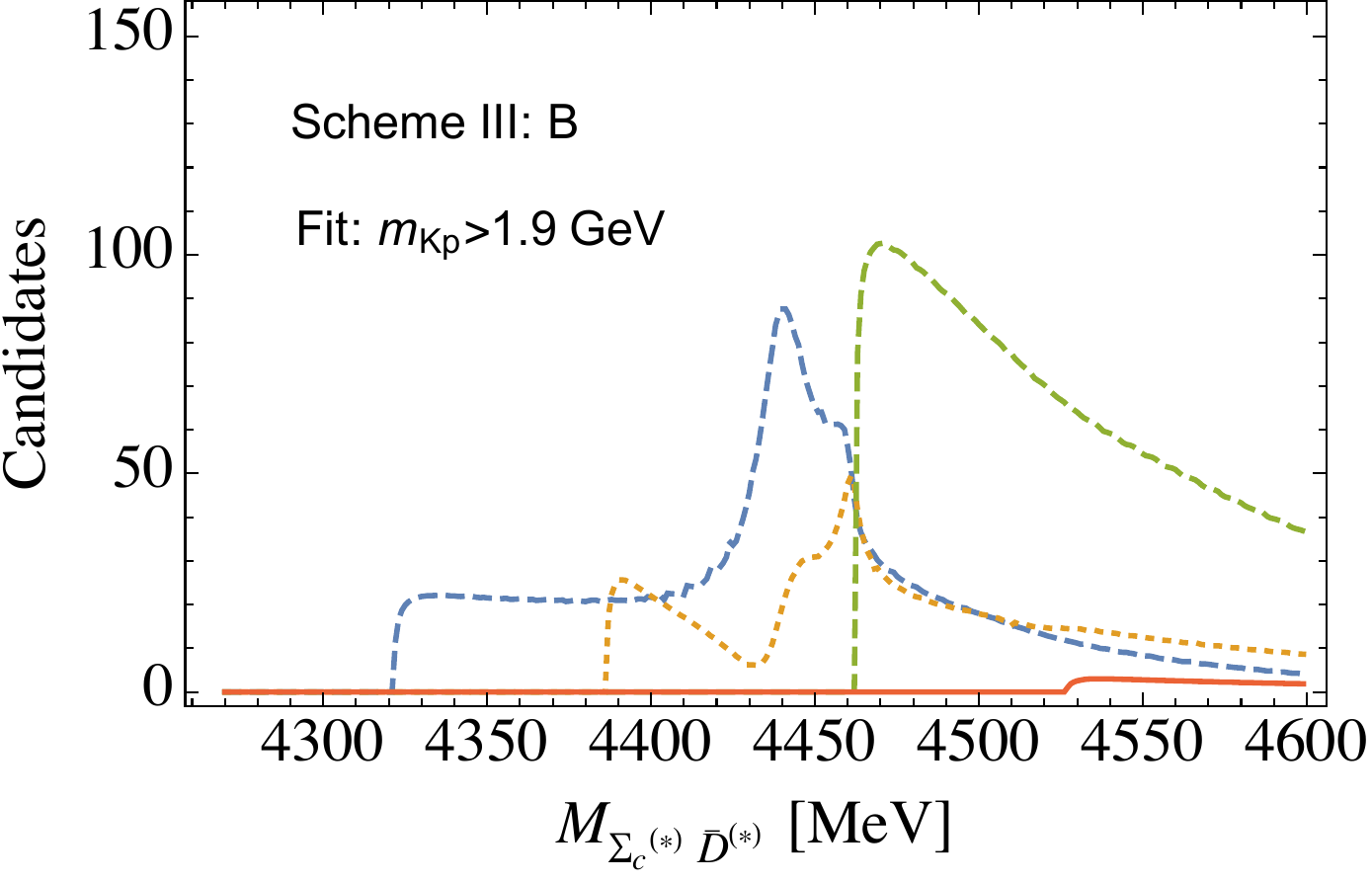}\\
  \includegraphics[width=0.32\textwidth]{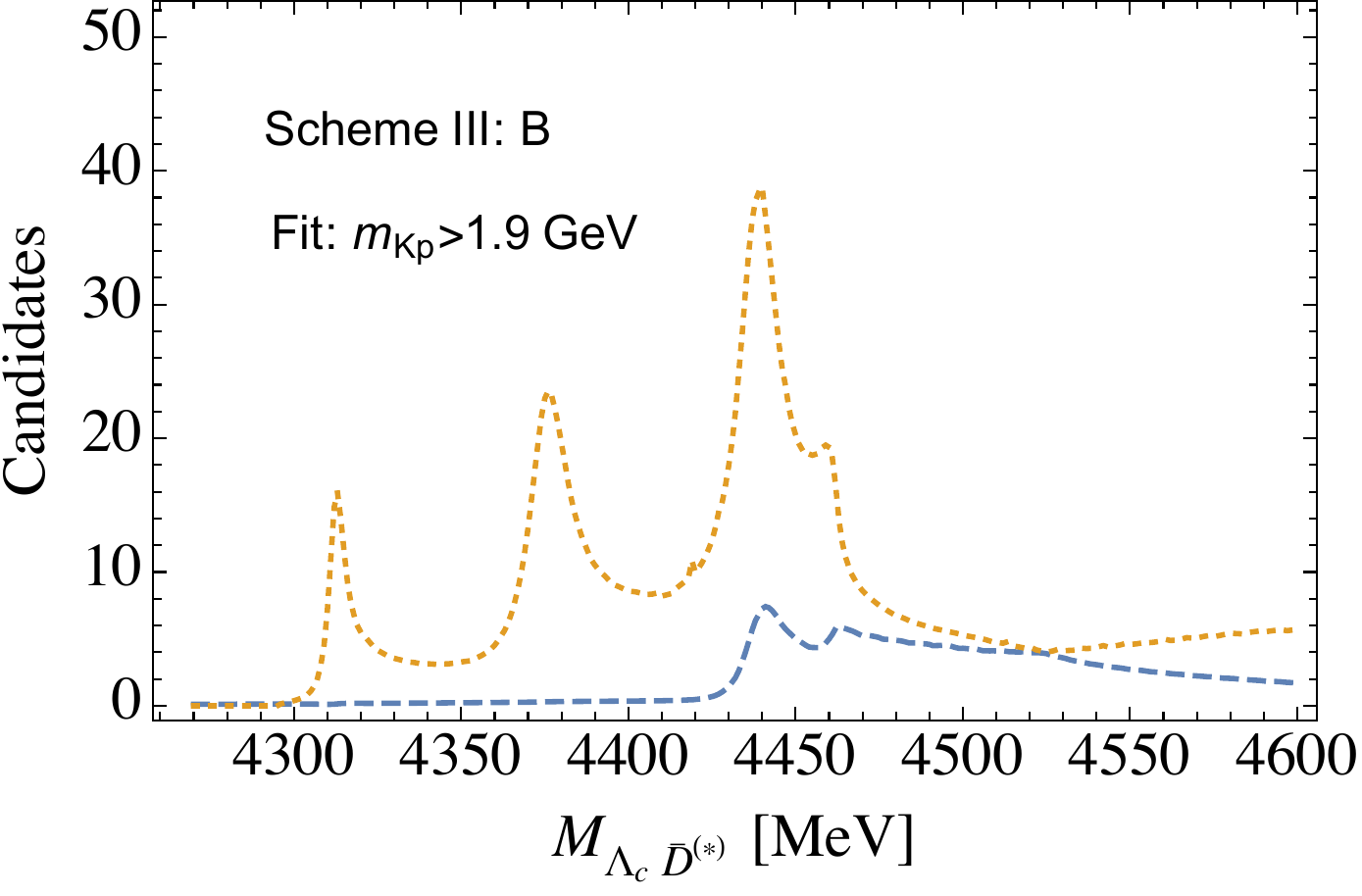}
  \includegraphics[width=0.34\textwidth]{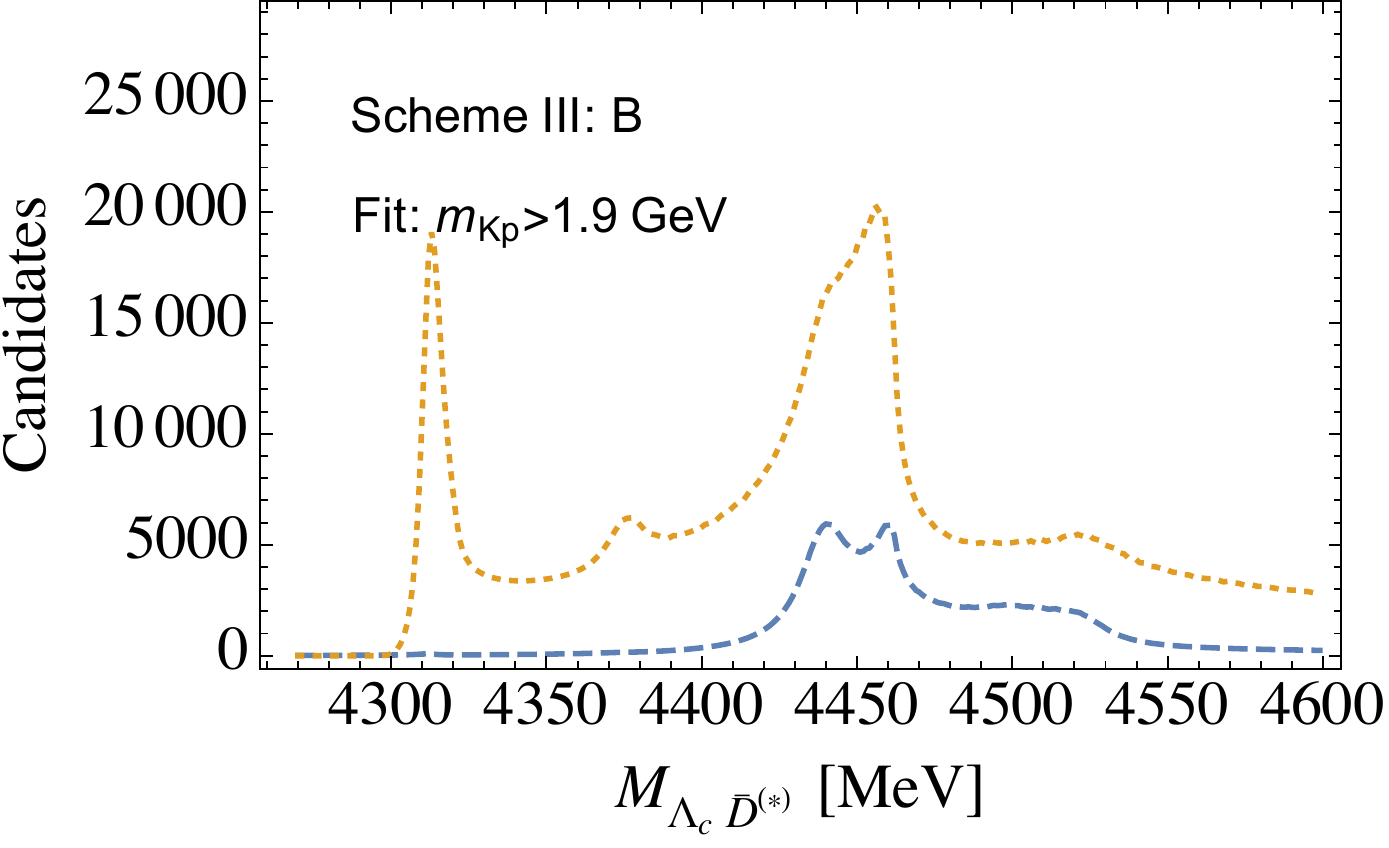}
  \includegraphics[width=0.32\textwidth]{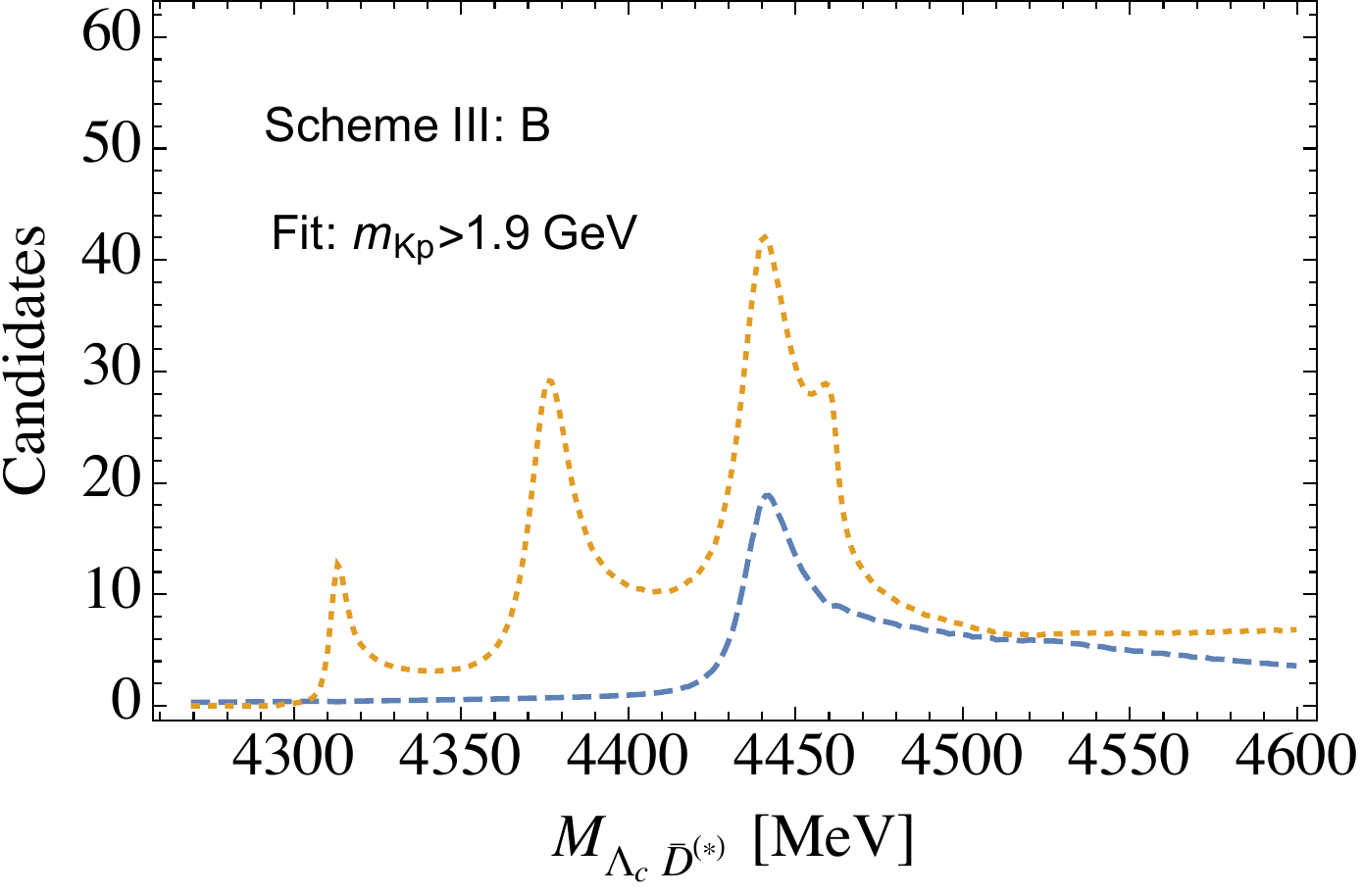}
  \caption{  \label{fig:III:sigh}
  Predictions for the line shapes in the $\sigh$ and $\lamh$ channels based on the fit results of scheme~III  (solution $B$) for three different backgrounds. The left, middle and right columns 
  show the results that are related to the blue, orange and green lines in Fig.~\ref{fig:III:fits}, respectively.}
\end{figure*}

As discussed above, the $\jp$ invariant mass distributions can be well described without including the $\lamh$ channels. It is found that the explicit inclusion of the $\lamh$ as well as the associated $\lamh\pi$ three-body cuts, denoted as Scheme III, on one hand does not improve the fit quality. On the other hand, the fitted inelastic parameters are not effectively constrained as they are very sensitive to the data sets used in the fits and to the background employed. It is hard to discriminate the contributions from the $\jp$ ($\etacn$) and $\lamh$ to the width of the $\pc$ states from the $\jp$ mass distribution alone. To illustrate it more clearly, we fits to the data with $m_{Kp}>1.9$ GeV with three different backgrounds, as shown in Fig.~\ref{fig:III:fits}. All the fits have values for $\chi^2$ very close to the best fit result. Nevertheless, the fitted parameters are very different, which leads to very different line shapes for the $\sigh$ and $\lamh$ productions, as shown in Fig.~\ref{fig:III:sigh}. In other words, the $\jp$ data alone are not enough to constrain the $\lamh$ interactions. Nevertheless, the predicted line shape for the $\etacn$ distribution is consistent with those of Scheme II, see i.e. in Fig.~\ref{fig:etacn}, as it is directly related to the $\jp$ channel via HQSS.

%
%

\section*{Acknowledgements}
This talk is based on the collaboration with Vadim Baru, Feng-Kun Guo, Christoph Hanhart, Ulf-G. Mei{\ss}ner, Jos\'e A. Oller, and Qian Wang.
The project is supported in part by the National Natural Science Foundation of China (NSFC) and  the
Deutsche Forschungsgemeinschaft (DFG) through the funds provided to the Sino-German Collaborative Research Center TRR110 ``Symmetries and the Emergence of Structure in QCD'' (NSFC Grant No. 12070131001, DFG Project-ID 196253076),
by the NSFC under Grants No. 11835015, No. 12047503, No. 11961141012, and No.12035007, by the Chinese Academy of Sciences (CAS) under Grants
No. QYZDB-SSW-SYS013 and No. XDB34030000, and by the Munich Institute for Astro- and Particle Physics (MIAPP) of the DFG cluster of excellence ``Origin and Structure of the Universe'', by the Spanish Ministry of Science and Innovation (MICINN) (Project PID2020-112777GB-I00), by the EU Horizon
2020 research and innovation programme, STRONG-2020 project, under grant agreement No.~824093, by Generalitat
Valenciana under contract PROMETEO/2020/023. 

%

\end{document}


<div id='footer'><table width='100